\documentstyle[aps,multicol]{revtex}
\input epsf

\newcommand{\be}{\begin{equation}}
\newcommand{\ee}{\end{equation}}
\newcommand{\ba}{\begin{eqnarray}}
\newcommand{\ea}{\end{eqnarray}}

\def\cP{{\cal P}} \def\bP{{\bf P}}  \def\var{{\rm var}}

\begin{document}
\draft

\title{Fine structure and complex exponents in \\
       power law distributions from random maps}
\author{Per J\"ogi$^{1,2,3}$ and Didier Sornette$^{2,3}$}
\address{
   $^1$ Department of Physics,
        University of California, Los Angeles, California 90095-1567 \\
   $^2$ Institute of Geophysics and Planetary Physics \\and
        Department of Earth and Space Sciences\\
        University of California, Los Angeles, California 90095-1567 \\
   $^3$ Laboratoire de Physique de la Mati\`{e}re Condens\'{e}e\\
   		CNRS and Universit\'{e} de Nice-Sophia Antipolis, Parc Valrose,
        06108 Nice, France 
}
\author{Michael Blank}
\address{
        Russian Academy of Sciences, \\
        Inst. for Information Transmission Problems, B.Karetnij Per. 19,
        101447, Moscow, Russia.
}

\maketitle

\begin{abstract} 
Discrete scale invariance (DSI) has recently been documented in
time-to-failure rupture, earthquake processes and financial crashes, 
in the fractal geometry of growth processes and in random systems. 
The main signature of DSI is the presence of log-periodic oscillations 
correcting the usual power laws, corresponding to complex exponents.  
Log-periodic structures are important because they reveal the presence 
of preferred scaling ratios of the underlying physical processes.  
Here, we present new evidence of log-periodicity overlaying the leading 
power law behavior of probability density distributions of affine random 
maps with parametric noise. The log-periodicity is due to intermittent 
amplifying multiplicative events. We quantify precisely the progressive 
smoothing of the log-periodic structures as the randomness increases 
and find a large robustness. Our results provide useful markers for 
the search of log-periodicity in numerical and experimental data.


\end{abstract}


\begin{multicols}{2}
\narrowtext

\section{INTRODUCTION}
\label{sec:intro}

Complex critical exponents and complex fractal dimensions have until 
recently been discussed only for hierarchical systems, be they man-made 
\cite{bessis,DIL,doucot,fournier,Vladmackey,Meurice,Newman,Stanley} or
naturally occurring as in the mammalian bronchial tree \cite{lung,West}.
These hierarchical systems are characterized by {\it discrete} scale 
invariance (DSI), a notion qualitatively similar to the concept of 
``lacunarity''. A signature of this DSI is the presence of log-periodic 
oscillations correcting the usual power laws, corresponding to  
{\em complex} exponents.

Recently, their occurrence in irreversible rupture
\cite{Anifrani,SorSam,SSS,Joh,Varnes,Ouillon} and growth processes
\cite{DLA,needles} as well as prior to financial crashes 
\cite{crashsor,Freund} have been documented. It has been suggested
\cite{SS} that complex exponents are rather common and should be 
looked for generically in any model whose critical properties are 
described by an underlying non-unitary field theory.  This excludes 
the usual homogeneous spin systems  in which the renormalization flow 
is a gradient \cite{Wallace}.  This includes models with non-local 
properties such as percolation and animals \cite{SS}, polymers and 
their generalizations, models of irreversible growth processes such 
as rupture \cite{Anifrani,SorSam,SSS,Joh,Varnes,Ouillon},
Diffusion-Limited-Aggregation (DLA) \cite{DLA,needles}, and models with 
quenched disorder like spin-glasses 
\cite{spinglass1,spinglass2,Khmel,Boya,Weinrib}. See \cite{review} for
a review.

Three outstanding problems remain\,:
\begin{itemize}
\item do we know all the physical mechanisms that can produce complex
      critical exponents?
\item how strong are the log-periodic structures and how robust are they
      with respect to noise and disorder?
\item does there exist a smooth invariant probability distribution
      (having a density), or is it discrete?
\end{itemize}
With respect to the first question, six situations have been discussed\,:
\begin{enumerate}
\item the presence of a built-in geometrical hierarchy
      \cite{bessis,DIL,doucot,fournier,Vladmackey,Meurice,Newman,Stanley,West}\,;
\item the diffusion in anisotropic quenched random lattices in which the
      hierarchy is constructed dynamically due to the probabilistic 
      encounters with traps \cite{BS}\,;
\item intermittent amplification processes \cite{Vlad}\,;
\item cascades of ultra-violet instabilities as in rupture and growth
      processes \cite{DLA,needles}\,;
\item non-local geometry \cite{SS}\,;
\item quenched disordered systems
      \cite{spinglass1,spinglass2,Khmel,Boya,Weinrib}.
\end{enumerate}

In regards to the second question, log-periodic oscillations of spin
systems on a fractal amounts to exceedingly small effects, typically 
of the order of $10^{-5}$ in relative value \cite{DIL,Meurice}. In 
contrast, it is still not fully understood why log-periodic structures 
seem to be many times stronger, of the order of $10\%$ or so, in rupture 
and growth processes. In addition, log-periodicity implies a prefered 
scaling ratio which, in nature, should be largely perturbed by disorder. 
Theoretical estimates of the effect of disorder on the log-periodic 
corrections indicate that they should be generally robust \cite{SS}. 
An important practical question is how much disorder or noise will make 
the log-periodic corrections too small to be observed. Ensemble averaging 
is also an issue as finite-size effects cause significant variations 
in the phase of the log-periodic oscillations.  Averaging may cause them 
to disappear.  This was observed in DLA clusters \cite{DLA} in which 
single cluster analysis uncovered the log-periodic structures while 
averaging procedures destroyed them.

In order to address these questions on the effect of disorder, we study a
simple, positive, random map with parametric noise\,,
\be
   X_{t+1} = a_t X_t + b_t,\;\;\mbox{with}\;a_t, b_t>0 \;.
   \label{fond}
\ee
The growth rate $a_t$ and the additional term $b_t$ are assumed to be 
pairs of positive identically distributed random values with the joint 
distribution function $\cP_{a,b}$.  In most of the cases treated below, 
we assume that $a_t$ and $b_t$ are independent, which yields 
$\cP_{a,b} = \cP_a\cP_b$.

\begin{figure}
  \centerline{\mbox{\epsfbox[0 0 260 182]{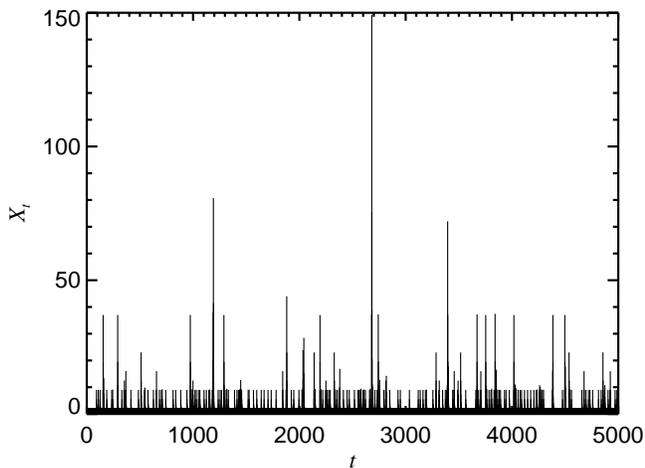}}}
  \caption{
           $X_t$ history for $a_t$ with a two point distribution 
           (section~\ref{sec:twop}) at $a=2$, $\xi=2$, $p=0.95$, 
           and $b_t=1$.
          }
  \label{fig1}
\end{figure}

It may seem that the linear model~(\ref{fond}) is so simple that it 
does not require a careful mathematical investigation.  This is 
however not the case, as the rather extensive mathematical analysis 
of the problem in \cite{Kesten} indicates. We will show a very unusual 
behavior of solutions of the difference equation~(\ref{fond}). It is 
known \cite{Kesten} that, provided some regularity assumptions hold 
and when the average rate of growth $\langle \ln a_t \rangle$ is 
negative, then the time series $X_t$ is stationary.  Furthermore 
$X_t$ is characterized statistically by a probability distribution 
function with a power law tail\,:
\be
   \cP_X(x) \sim x^{-(1+\mu)}\;,
   \label{power}
\ee
when the equation for $\mu$,
\be
   \langle a^{\mu} \rangle = 1
   \label{mueq}
\ee
has a positive solution \cite{Calan,Haan,Kesten,SC,Solomon,SK}.

The power law distribution function stems from an intermittent and
transient  amplification occurring when several successive $a_t$ 
are larger than $1$.  Its origin is thus in the class of intermittent 
amplifications \cite{Vlad} and intermittent trapping \cite{BS} 
mechanisms. We may therefore expect complex valued $\mu$ exponents.  
This should lead to the occurrence of detectable log-periodic 
corrections in the leading simple power law behavior.

We are here concerned with the continuity or discreteness of this
distribution function and with the strength and detection of the 
potential log-periodic corrections, especially as a function of the 
distributions $\cP_a$ and $\cP_b$. Intuitively, the broader these 
distributions are, the weaker we expect the log-periodic corrections 
to be since a log-periodicity is the signature of a favored scaling 
ratio. This preference must disappear as the disorder increases. We 
aim to carefully quantify this scenario, for the benefit of future 
analysis of log-periodicity.

\begin{figure}
  \centerline{\mbox{\epsfbox[0 0 260 182]{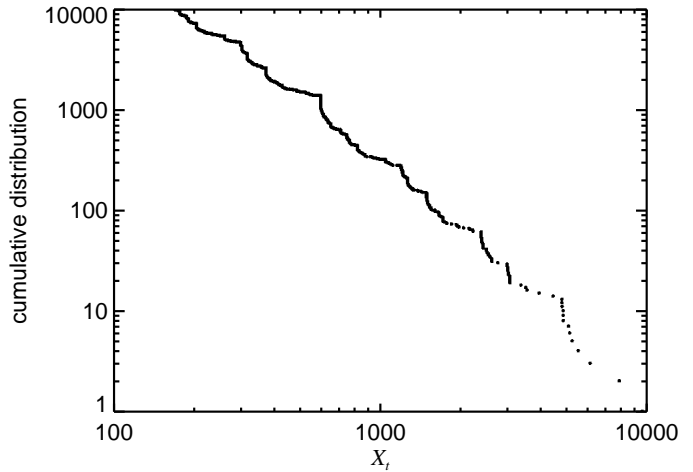}}}
  \caption{
           Cumulative distribution of the $10^4$ largest iterates 
           among $10^8$ realized for $a_t$ with a two point 
           distribution (section~\ref{sec:twop}) at $a=2$, $p=0.95$, 
           and $b_t=1$.
          }
  \label{fig2}
\end{figure} 

In the next two sections, we recall useful information, discuss a 
connection with products of random matrices and iterated function 
systems and review the so-called transition operator approach 
determining the probability density function (pdf) $P(X)$.  We 
then discuss the case in which $a_t$ take only two values ${1/a}$ 
and $a^{\xi}$ with probability $p$ and $1-p$ respectively, where 
$a>1$ and $\xi > 0$, first in the case of a fixed $b=1$ and then 
with increasingly widening $\cP_b$ distributions. We then analyze 
the case in which $a_t$ is broadly distributed and discuss the 
detection criteria for the log-periodic corrections.

In addition to the present focus as a paradigm of system exhibiting 
complex exponents, this random map (\ref{fond}) has been introduced 
in various contexts, for instance in the physical modelling of 1D 
disordered systems \cite{Calan} and the statistical representation 
of financial time series \cite{Haan}.  The variable $X_t$ is known 
in probability theory as a Kesten variable \cite{Kesten}.  The map 
(\ref{fond}) describes for instance the time evolution of a fish 
population $X_t$ with $a_t$ depending on the rate of reproduction 
and on the depletion rate due to fishing as well as environmental 
conditions and $b_t$ describing the input due to restocking from an 
external source such as a fish hatchery, or from migration from 
adjoining reservoirs \cite{SK}.  The random map (\ref{fond}) can 
also be applied to other problems of population dynamics, epidemics, 
investment portfolio growth, and immigration across national borders 
\cite{SK}.  Variations of this model have recently been proposed for 
the analysis of crop control in the presence of weed infestion
\cite{Hugues}. Models of enonomic evolution typically involve a system 
of affine coupled equations of the type (\ref{defaffine}) below, which 
are multi-dimensional generalization of (\ref{fond}).  For instance, 
the economic model of Keynes in its simplest form links consumption, 
investment and production in a linear affine system of deterministic 
equations.  The system (\ref{defaffine}) corresponds to a generalization 
in which the coefficients of the auto-regression are allowed to 
fluctuate in time to account for uncertainty.  More generally, models 
used in econometrics \cite{Greene} are very similar to (\ref{fond}) 
and  (\ref{defaffine}) below, even if they usually assume constant
coefficients.

It is probably true that (\ref{fond}) is one of the simplest 
{\it linear} stochastic equation that can provide an alternative 
modeling strategy for describing complex time series. We note that 
a nonlinear version with a quadratic nonlinearity (corresponding to 
the logistic equation with random multiplicative noise) has recently 
been shown to lead to a new type of crisis, in that there is a 
sudden qualitative change in the chaotic dynamical behavior induced 
by variations of the parameters \cite{Yang}.  We do not discuss 
these properties but restrict our considerations to the affine random
map (\ref{fond}).

\section{RESULTS ON THE KESTEN AFFINE RANDOM MAP}
\label{sec:resu}

\subsection{Formal solution}
\label{subsec:form}

The formal solution of (\ref{fond}) for $N \ge 1$ reads
\be
   X_{t+N} = (\prod_{l=0}^{N-1} a_{t+l}) X_t + \sum_{l=0}^{N-1} 
   b_{t+l} \prod_{m=l+1}^{N-1} a_{t+m} \;,
   \label{azjhg}
\ee
where we define $\prod_{m=N}^{N-1} a_{t+m} \equiv 1$ for the special
value $l=N-1$.  It is clear that the $\prod_{l=0}^{N-1}a_{t+l}$ 
multipliers of (\ref{azjhg}) control the $X_t$ dynamics.  Thus,
$X_{t+N}$ diverges (resp. remains bounded) if the average logarithmic 
growth factor $\langle \ln a_t \rangle$ is positive (resp. negative). 
Here, we focus our attention on the case
\be
   \langle \ln a_t \rangle < 0 \;.
   \label{eq:lna0}
\ee
In this regime, we notice the role of $b_t$ which provides a 
{\it reinjection} mechanism \cite{SC} allowing $X_t$ to fluctuate 
without converging to zero, as it would if $b_t$ vanished.

\subsection{Product of random matrices}
\label{subsec:prod}

The map (\ref{fond}) can be written as a product of random 
$2 \times 2$ matrices\,:
\be
   \left(
         \begin{array}{c}
                X_{t+1}\\ 1
         \end{array}
   \right) =
   \left(
         \begin{array}{cc}
                a_t  &  b_t\\ 0 & 1
         \end{array}
   \right)
   \left(
         \begin{array}{c}
                X_{t}\\ 1
         \end{array}
   \right) \;.
   \label{matr.repr}
\ee
By Furstenberg's theorem, the norm $||V_t||$ ($\sim X_t$ for 
large $X_t$) of the $t$-th vector
\be
   V_t \equiv \left( \begin{array}{c} X_{t}\\ 1 \end{array} \right)
   \label{furst.them}
\ee
grows as \cite{Paladin}
\be
   ||V_t|| = ||V_0|| e^{\lambda_1 t} \;,
   \label{grow.pala}
\ee
where $\lambda_1$ is the largest Lyapunov exponent of the product 
of the random matrices.  The $2 \times 2$ matrices are triangular 
and thus
\be
   \lambda_1 = \max \{ \langle \ln a_t \rangle, \,0  \} \;.
   \label{lyap.tria}
\ee
We recover the exponential growth regime of $X_t$ for
$\langle \ln a_t \rangle > 0$.  In the reverse case,
$\langle \ln a_t \rangle < 0$, the Lyapunov exponent 
is zero, which corresponds to the marginal case between 
exponential growth and exponential decay.  This is the 
regime where one usually encounters power law behavior, 
for instance in power law sensitivity to initial conditions 
in dynamical systems at the onset of chaos \cite{Tsallis}.

It is worth noticing that this zero Lyapunov exponent is 
different from the directly measured Lyapunov exponent of 
(\ref{fond}).  Indeed, the solution (\ref{azjhg}) shows that 
a perturbation $\delta X_t$ at time $t$ gives an error
$\delta X_{t+N} = \delta X_t \prod_{l=1}^{N-1} a_{t+l}
\sim e^{(N-1) \langle \ln a \rangle}$.  This corresponds to a 
{\it negative} Lyapunov exponent for the case studied here 
(\ref{eq:lna0}), equal to $\langle \ln a \rangle$.  This would 
lead one to conclude that the dynamics is trivial.   Generally 
speaking, the widespread opinion in the physical community that 
notions like chaos, etc., have some strict correspondence to the 
positivity of Lyapunov exponents is not quite correct.  See, 
for example, the detailed discusssion of nonchaotic dynamical 
systems  with positive Lyapunov exponents and vice versa, and
futher references in  \cite{Bl,BBF}. Here, the usual calculation 
of the Lyapunov exponent is not sensitive to the ``reinjection'' 
mechanism brought by the $b_t$ term. By construction, the matrix 
formulation (\ref{matr.repr}) takes this effect into account. 
The resulting vanishing Lyapunov exponent alerts us to the 
possibility of complex behavior.

\subsection{Iterated Function System}
\label{subsec:iter}

We also mention the relationship with Iterated Function Systems 
(IFS), which are defined as follows \cite{Barnsley}. One first 
defines an affine transformation $W$ from $R^D$ to $R^D$:
\be
   W[{\bf x}] = {\bf A x} + {\bf b} \;,
   \label{defaffine}
\ee
where ${\bf A}$ is a $D \times D$ matrix and ${\bf b}$ a vector 
in $R^D$.  An affine transformation is contractive if there exists 
a Lipschitz constant $s < 1$ such that
\be
   |W[{\bf x}] - W[{\bf y}]| < s |{\bf x} - {\bf y}| \;.
   \label{lipsc.cnst}
\ee
An IFS consists of $N$ affine transformations $W_i$ and a set 
of probabilities $p_i > 0$ with $\sum_{i=1}^N p_i= 1$.  Starting 
with a given set of points, the IFS code consists in applying 
to it an infinite sequence of transformations, each of them 
being chosen with its corresponding probability.  In general, IFS
codes satisfy the average contractive condition\,:
\be
   s_1^{p_1}s_2^{p_2}\cdots s_N^{p_N} < 1 \;.
   \label{contract}
\ee
Taking $D=1$, we see that (\ref{defaffine}) is the same as (\ref{fond}),
where $N$ is the number of different values taken by $a_t$ (suppose for
simplicity that $b_t$ is constant) with their respective probabilities 
$p_i$.  In other words, the affine random map (\ref{fond}) is a 
one-dimensional IFS.  Then, the Lipschitz constant $s_i$ is equal to 
the $i$-th value $a_i$ that $a_t$ can take. Condition (\ref{contract}) 
then becomes the familiar
\be
   \sum_{i=1}^N p_i \ln a_i \equiv \langle \ln a \rangle < 0 \;.
   \label{lipsc.ctrct}
\ee
This retrieves the regime (\ref{eq:lna0}) discussed above.  Usually, 
IFS are studied in situations where all the affine transformations 
have their Lipschitz constant individually negative, {\it i.e.} all 
are contractive.  The present work (where $D=1$) deals with a rather 
special but very interesting situation where some of them are dilating, 
while on average the set of transformations is contractive. This 
correspondence and the discovery that power law distributions are 
found when some of the transformations of the IFS are dilating 
suggests to us an investigation of the behavior of similar intermittent 
dilating IFS in higher dimensions, where rotations are added to the 
translation and dilation processes. This is left for future work.  Here
we will next use the correspondence with IFS to understand intuitively 
the fractal structures found when the $a_t$ take a finite number of values.

\subsection{Probability density function}
\label{subsec:prob}

Calling $P_{a_t}$, $P_{b_t}$, and $P_{X_{t+1}}$ the pdf's of $a_t$, 
$b_t$, and $X_{t+1}$ respectively (and assuming that they are 
integrable functions), then the pdf of $X_t$ (obtained by the 
standard Markov argument) obeys the following equation\,:
\ba
   P_{X_{t+1}}(X)&=&\int_{-\infty}^{\infty} P_{a_t}(a) d{a}
      \int_{-\infty}^{\infty} P_{b_t}(b) d{b} \nonumber \\
                 & & \mbox{}\times \int_{-\infty}^{\infty} 
      P_{X_t}(Y) \delta(X - a Y - b)d{Y} \;,
   \label{eq:pdf0}
\ea
or
\be
   P_{X_{t+1}}(X)=\int_{-\infty}^{\infty} {P_{a_t}(a) \over a} d{a}
      \int_{-\infty}^{\infty} P_{b_t}(b)
      P_{X_t}({{X-b}\over{a}}) d{b} \;.
   \label{eq:pdf1}
\ee
The two pdfs $P_{X_t}$ and $P_{X_{t+1}}$ approach a common stationary 
pdf, $P(X)$, for large $t$ \cite{SC,Solomon}.  We  are interested in 
the description of the tail of $P(X)$, {\it i.e.} for $X \gg b$.  We 
can then neglect the $b$ term of $P_{X_t}((X-b)/{a})$ in the r.h.s. 
of (\ref{eq:pdf1}).  This allows us to simplify (\ref{eq:pdf1}) into
\be
   P(X)=\int_{-\infty}^{\infty} {P_{a_t}(a) \over a} P({X \over a}) d{a}
   \quad\mbox{for large}\quad X \;,
   \label{eq:pdf2}
\ee
using $\int_{-\infty}^{\infty} P_{b_t}(b_t) d{b_t}=1$.  Since 
(\ref{eq:pdf2}) is linear in $P(X)$, the general solution can 
be written as a sum over a set of particular solutions \cite{Calan}. 
These solutions are composed of power laws and faster decaying 
functions (exponentials).  The set of power law solutions is 
obtained by assuming the form $P(X)\sim X^{-(1+\mu)}$ for $X \gg 1$.  
This yields (\ref{mueq}) determining the exponent $\mu$.

The inequality (\ref{eq:lna0}) and the equation (\ref{mueq}) are 
the corner stones of our analysis. We construct and analyze several 
examples whose parameters are constrained by (\ref{eq:lna0}) and we 
study the solutions of (\ref{mueq}) and compare them to direct 
numerical simulations.

\section{TRANSITION OPERATOR APPROACH}
\label{sec:trans}

One of the obstacles in the implementation of the approach in the
previous section is that we need to assume that all the considered
distributions have densities, which is not the case with at least
some of our examples. To study a more general situation, let us
consider the so-called transition operator approach.

\subsection{Transition operator approach and non-smooth distributions}
\label{subsec:trans}

According to our definitions, (\ref{fond}) defines a Markov chain.  
We can define the transition operator $\bP$ of this random process as\,:
\be 
   \bP h(x) = \int h(\frac{x-b}{a}) \, d\cP_{a,b}(a,b)
   \label{tr.operator}
\ee
for any integrable function $h$. This operator describes the image 
of a distribution density under the action of our random process. 
If an invariant probability density $P$ exists it should satisfy
\be
   \bP P = P \;.
   \label{ME}
\ee
The introduction of (\ref{tr.operator}) is justified by the fact that 
this integral operator allows for the study of non-smooth and even 
discontinous distributions.

Consider a simple implementation of a random selection scheme.
Assume that $0 < a_1 < 1 < a_2$, $0 < b_1$, $0 < b_2$ for the two maps\,:
\ba
   x &\to& a_1x + b_1 \,
     \label{eq:bm01}\\
   x &\to& a_2x + b_2 \,
     \label{eq:bm02}
\ea
such that (\ref{eq:bm01}) is chosen with the probability $p$, 
while (\ref{eq:bm02}) is chosen with the probability $1-p$. 
The corresponding joint distribution $\cP_{a,b}$ for the random 
variables $a_t$ and $b_t$ is singular.  Furthermore, these two 
random variables heavily depend on each other.  Therefore 
(\ref{eq:pdf1}) cannot directly be used.  However it is trivial 
to specify the corresponding transition operator\,:
\be
   \bP h(x) = \frac{p}{a_1} h(\frac{x-b_1}{a_1})
              + \frac{1-p}{a_2} h(\frac{x-b_2}{a_2}) \;.
   \label{eq:bm03}
\ee
Moreover, this description is easily generalized to the case with an 
arbitrary (albeit finite) number of linear maps $x \to a_ix + b_i$, 
where each is selected with the probability $p_i$ (provided $\sum_i p_i=1$),
\be
   \bP h(x) = \sum_i \frac{p_i}{a_i} h(\frac{x-b_i}{a_i}) \;.
   \label{main.eq}
\ee

The representation~(\ref{main.eq}) shows that our random system lacks a
smooth (and even bounded) invariant density for any choice of the positive
coefficients $a_i$ and $b_i$. Assume on the contrary that such pdf $h(x)$
exists. This has to satisfy $\bP h(x) = h(x)$ for any $x \in [0,1]$.  To
proceed further, we need to estimate the variation of the image of the $h$
function. The variation of a function is, roughly speaking, an integral
of the modulus of the derivative of the function over its domain.  In the
case of monotonic functions, it can be shown that
\be
   \var(\bP h) = \var(h) \sum_i \frac{p_i}{a_i} \;.
   \label{eq:bm04}
\ee
According to (\ref{contract}), we note that
\be
   \prod_i a_i^{p_i} < 1 \;,
   \label{eq:bm05}
\ee
which yields
\be
   \sum_i \frac{p_i}{a_i} > 1\;.
   \label{eq:bm06}
\ee
Therefore, each time we apply the transition operator, the variation of the
image of a function is multiplied by the factor $\sum_i \frac{p_i}{a_i}>1$.
Hence the limit distribution (if it exists) cannot be a function of bounded
variation.

\begin{figure}
  \centerline{\mbox{\epsfbox[0 0 260 182]{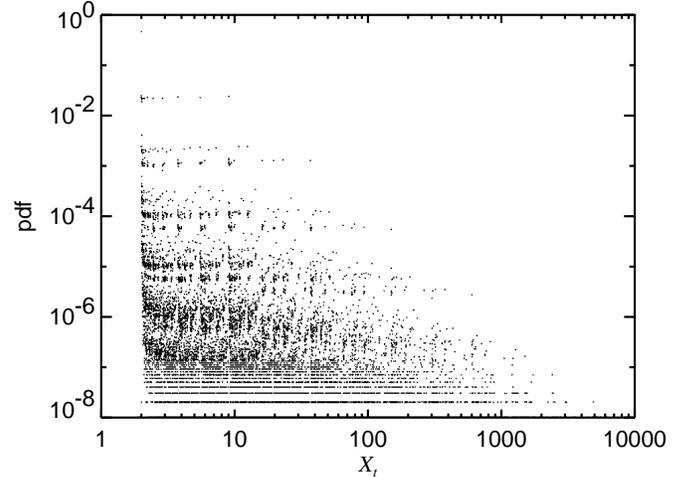}}}
  \caption{
           The (numerically obtained) pdf for $a_t$ with a two point 
           distribution (section~\ref{sec:twop}) at $a=2$, $\xi=2$, 
           $p=0.95$ and $b_t=1$ ($10^8$ iterates, $10^4$ equispaced 
           bins per unit of $\log X_t$).
          }
  \label{fig3}
\end{figure}

A special case was treated in \cite{Calan} with a finite system of 
random maps producing a discrete invariant distribution\,:
\ba
   x &\to& 1 \qquad\hbox{with probability}\quad p \;,
     \label{eq:bm07}\\
   x &\to& ax + 1 \qquad\hbox{with probability}\quad 1-p \;. 
     \label{eq:bm08}
\ea
In this case, it is easy to find the invariant distribution analytically.
However, since the first map has a zero value of the multiplier $a$, the
system does not satify our assumptions that all coefficients should be
positive.

\subsection{Existence of pdf for the case of the smooth distribution of
            coefficients}
\label{subsec:exis}

In this subsection, we study a more general case, where we have a random
map $x \to ax + b$ with random coefficients $a,b$, whose joint probability
distribution is $\cP(a,b)$. The case considered above corresponds to
the discrete distribution $\cP(a,b)$. Our main aim here is to prove that,
if the distribution $\cP(a,b)$ has a density $p(a,b)$ with ``good enough''
properties, then the random map system also has a finite invariant density.
Indeed, consider the corresponding transition operator:
\be
   \bP h(x) = \int\int \frac{p(a,b)}{a} h(\frac{x-b}a)\,da\,db \;.
    \label{eq:bm09}
\ee
After the change of variables $\xi=x-b$ this operator becomes
\be
   \bP h(x) = - \int\int \frac{p(a,x-\xi)}{a} h(\frac{\xi}a) \, da \, d\xi \;.
   \label{eq:bm10}
\ee
Assume now that
\be
   \var_b p(a,b) \le C < \infty
   \label{eq:bm11}
\ee
for any $a>0$, where $\var_b p(a,b)$ stands the variation of 
the $p(a,b)$ function with respect to the second variable. 
Using the above representation, we find that
\ba
   \var(\bP h) &\le& \int\int \frac{C}{a} h(\frac{\xi}a) \, da \, d\xi
        = C \int \left( \int h(\frac{\xi}a) \, d(\frac{\xi}a) \right) \, da 
        \nonumber\\
               &=& C (\int h(x) \, dx)\int da \;.
        \label{eq:bm12}
\ea
Since $h$ is assumed to be the density of a probability distribution, then 
$\int h(x) \, dx$ is finite. This shows that the variation of $\bP h$ is 
universally bounded from above.  The existence of the invariant distribution 
\cite{Kesten} then proves the existence of the pdf.

\subsection{Markov dependent choice of the subsequent map}
\label{subsec:mark}

Our earlier discussion of random map systems assumed that the choice of
a subsequent map $x \to a_ix + b_i$ does not depend on the immediatly
antecedent map chosen. This is a relatively strong restriction and in this
section we shall show that this assumption is not necessary.  We will
show that, for the stationary process, the representation of the transition
operator depends only on the stationary probabilities of the random choices
$p_i$ and not on the transition probabilities  between subsequent maps.

Assume that currently the map $x \to a_ix + b_i$ was chosen, then the
conditional probability to choose the map $x \to a_jx +b_j$ is equal to
$p_{ij}$. The process of the random choice is governed by the finite state
Markov chain with the transition probabilities $(p_{ij})$.  Assume that this
Markov chain is ergodic and denote by $p_i$ its unique invariant distribution.
Then our entire system is still a Markov chain, whose transition operator is:
\ba
   &&\bP h(x) = \sum_i p_i \left(\sum_j \frac{p_{ij}}{a_j} h(\frac{x-b_j}{a_j})
      \right) \nonumber \\
   &=& \sum_j \frac1{a_j} h(\frac{x-b_j}{a_j}) \sum_i p_i p_{ij}
      = \sum_j \frac{p_j}{a_j} h(\frac{x-b_j}{a_j}) \;.
      \label{eq:bm13}
\ea
It depends only on the stationary probabilities $p_i$.  As a result, we
immediately see that all asymptotic properties also depend only on the
choice of $p_i$.

The generalization of our argument for the general case where the joint
distribution of the coefficients $a$ and $b$ may have both discrete and
continuous components is straightforward.

\section{TWO-POINT DISTRIBUTIONS}
\label{sec:twop}

Let us return now to the question about the asymptotic (as $x \to \infty$)
properties of pdf for the case of only two linear maps. The above derivations
show that these asymptotic properties do not depend on the choice of the
additional terms $b_i$ (as long as they are non-zero and well-behaved). Let
\be
   a > 1,\quad 0 < p < 1, \quad\mbox{and}\quad \xi>0 \;,
   \label{eq:ads2}
\ee
such that $a_1=1/a<1$ and $a_2=a^\xi>1$.  Equation (\ref{eq:pdf2}) becomes
\be
   P(X)=p a P(aX) + (1-p) a^{-\xi} P(a^{-\xi}X) \;.
   \label{eq:pft1}
\ee
Condition (\ref{eq:lna0}) imposes the requirement
\be
   {\xi \over 1+\xi} < p < 1 \;,
   \label{eq:pbxi}
\ee
and equation (\ref{mueq}) leads to
\ba
   (1-p) z^{1+\xi} - z + p &=& 0, \nonumber \\
   \qquad\mbox{where}\; z&\equiv & a^{\mu}\;\mbox{and $z$ complex} \;.
   \label{eq:zxip}
\ea

\subsection{$\xi$ integer}
\label{subsec:xint}

At first take $\xi = 1$, then from (\ref{eq:pbxi}), we see that $p$ 
must be within ${1 \over 2} < p < 1$.  The two {\it real} solutions 
of (\ref{eq:zxip}) are $z_{\pm} = {1 \pm \sqrt{\Delta} \over 2(1-p)}$, 
where $\Delta = 1 - 4 p (1-p) \geq 0$. From the definition $z = a^{\mu}$ 
(and $e^{i 2n\pi}=1$ for any integer n), we obtain
\be
   \mu_{\pm,n} \equiv \mu_{R} + i \mu_{I} =  {\ln z_{\pm} \over \ln a} + i
       {2\pi n \over \ln a} \;.
   \label{imagin}
\ee
The pdf of $X_t$ is thus of the form
\be
   P(X_t)=\sum_{\pm,n} {C_{\pm,n} \over X_t^{(1 + \mu_R)}}
          \cos(\mu_{I}\ln X_{t}) \;.
   \label{eq:cmuc}
\ee
The prefered scaling ratios are obtained by the factors of $X_t$ reproducing
the same values of the cosine, {\it i.e.} $a^{1 \over n}$, with $n$ integer.
The discrete scale invariance is simply the result of the intermittent
amplification by the fixed factor $a$. The log-periodicity is thus trivially
associated with the discrete multiplicative structure.

When $\xi = 2$, three {\it real} $z$ solutions exist for $z$, for the 
allowed range ${2 \over 3} < p <1$. The imaginary part of $\mu$ thus 
stems from the same technical reason as for $\xi = 1$ and reflects the 
intermittent amplification by the factor $a^2$.

In general if $\xi=N$ is a positive integer, (\ref{eq:zxip}) obeys
\be
   (1-p)z^{N+1}-z+p=0 \quad \Leftrightarrow \quad (z-1)Q_{N}(z;p)=0 \;,
   \label{eq:qpol}
\ee
where
\be
   Q_{N}(z;p)\equiv (1-p)\sum_{k=1}^{N} z^k-p \;.
   \label{eq:qdef}
\ee

The root structure of $Q_{N}(z;p)=0$ for $\Re z>0$ is determined with
e.g. Routh's algorithm~\cite{Gant}. It can be shown that for $N<5$, this
polynomial has only one root with $\Re z>0$ and that $\Im z=0$ for this
single root. However for $N\ge5$ there always exist roots such that
$\Re z>0$ and $\Im z\ne0$.

To illustrate the integer $\xi$ regime, Fig.~\ref{fig1} shows a segment
of the $X_t$ history for the case $a=2$, $\xi=2$, $p=0.95$, and a constant
$b_t=1$.  Most iterates are small while rare intermittent excursions 
explore very large values. The (numerically obtained) cumulative 
distribution is shown in the log-log plot of Fig.~\ref{fig2}.

\begin{figure}
  \centerline{\mbox{\epsfbox[0 0 270 307]{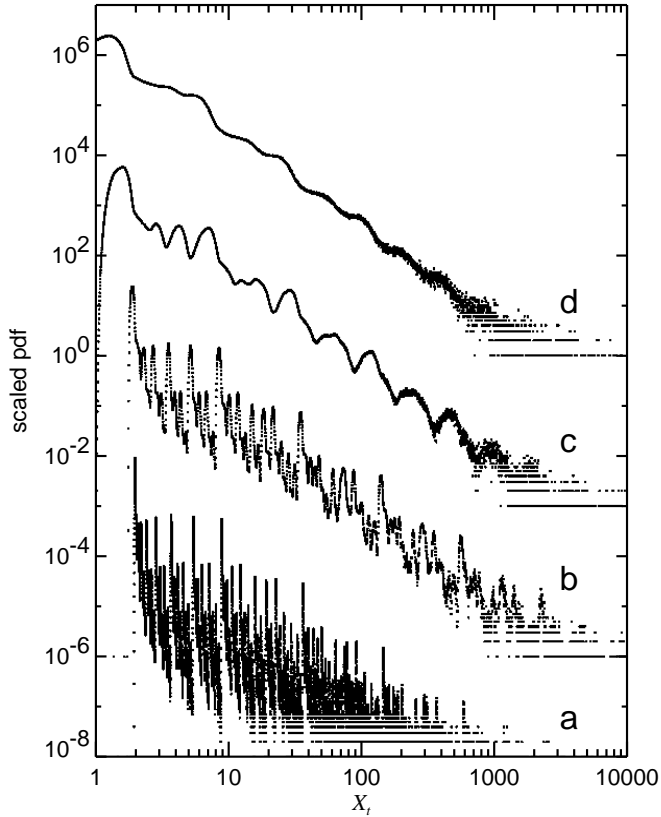}}}
  \caption{
           The scaled pdf of $X$ given by (\ref{fond}) for $a_t$ 
           with two point distribution at $a=2$, $\xi=2$, $p=0.95$, 
           and $b_t$ uniform with (a) $1\times{\rm pdf}$ and 
           $\beta={31\over32}$ ($10^8$ iterates, $10^4$ equispaced 
           bins per unit of $\log X_t$), (b) $10^3\times{\rm pdf}$ 
           and $\beta={7\over 8}$ ($10^9$ iterates, $10^3$ equispaced 
           bins per unit of $\log X_t$, the same for (c--d)), (c) 
           $10^6\times{\rm pdf}$ and $\beta={1\over2}$, (d) 
           $10^9\times{\rm pdf}$ and $\beta=0$.
          }
  \label{fig4}
\end{figure}

A complex structure, reminiscent of a devil staircase, overlays an 
average linear decay.  The structure corresponds to all possible 
values of $n$ in the imaginary part (\ref{imagin}) of the exponent 
$\mu$, where the largest $n$ provide the smallest details of the 
cumulative distribution.  Figure~\ref{fig3} shows the (numerically 
obtained) pdf, {\it i.e.} the derivative of Fig.~\ref{fig2}.  We 
observe a self-similar structure, as expected from the correspondence 
with the IFS discussed in section~\ref{subsec:iter} (IFS in general 
encode stochastic fractal structures \cite{Barnsley}).  We also 
observe that the distribution seems to be nowhere continuous, as 
expected from the derivation in section~\ref{subsec:trans}.

It is interesting to progressively coarse-grain this self-similar 
structure by introducing a disorder on $b_t$.  This is accomplished 
by chosing $b_t$ uniformly in the interval $[\beta, 1]$.  The value 
$\beta=1$ recovers the ordered case $b_t=1$.  Decreasing $\beta$ 
corresponds to increasing the disorder.  Figure~~\ref{fig4}  shows 
the pdf $P(X_t)$ for decreasing values $\beta={31 \over 32}, {7\over 8}, 
{1 \over 2}, 0$ (while keeping $a=2$, $\xi=2$, and $p=0.95$).  The 
roots of (\ref{eq:zxip}) are $z_{0}=1$ and $z_{\pm}={-1 \pm{\sqrt77}\over 2}$, 
or $\mu_{R0}=0$, $\mu_{I0}=n{2\pi\over \ln2}$, $\mu_{R+}\approx 1.9588$, 
$\mu_{I+}=\mu_{I0}$, $\mu_{R-}\approx 2.5890$, $\mu_{I-}=(1+2n) 
{\pi\over \ln2}$ for integer $n$.

\begin{figure}
  \centerline{\mbox{\epsfbox[0 0 260 369]{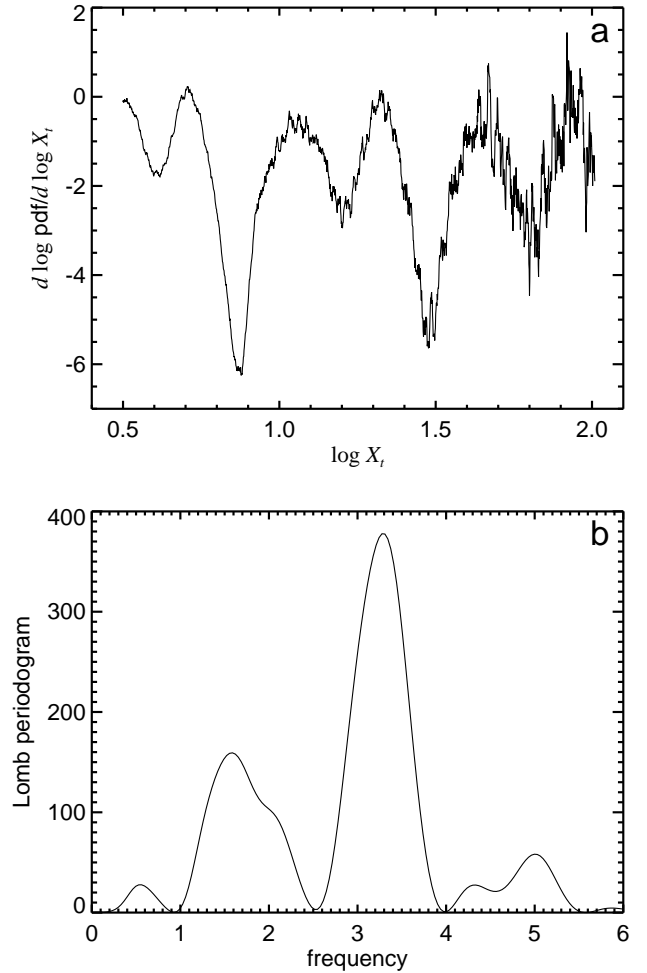}}}
  \caption{
           (a) The logarithmic derivative of a portion of the pdf 
           tail for the upper ($\beta=0$) trace in (\ref{fig4}). 
           (b) The Lomb periodogram of (a).
          }
  \label{fig5}
\end{figure}

The ``frequency'' of a log-periodic oscillation is defined by
\be
   f(n) \equiv {\mu_{I}(n) \over {2\pi \log e}} \equiv {1 \over \ln
       \lambda}\;,
   \label{eq:frqd}
\ee
where we define $\lambda$ as the scaling ratio associated with the
log-periodicity \cite{SSS,SS}.  Here $f(1)_{0}=f(1)_{+}\approx 3.3219$, 
$f(0)_{-}\approx 1.6609$, and $f(1)_{-}\approx 4.9828$.  These numbers 
are compared to the spectrum analysis of the tail portion of the pdf.  
We use the logarithmic derivative of the $\beta=0$ pdf of Fig.~\ref{fig4} 
to get a data with zero average slope (its averave value is the leading 
power law exponent).  This is represented in Fig.~\ref{fig5}~a. Its Lomb 
periodogram spectrum \cite{Numerci} is shown in Fig.~\ref{fig5}~b and 
yields four frequencies $0.6$, $1.6$, $3.3$, and $5.0$. The smallest of 
these is attributed to the inverse of the (log) tail length used.  The 
others are in good agreement with the predictions $f(0)_{-}$, $f(1)_{+}$,
$f(1)_{-}$ in ascending order. The small, partially hidden, bump at $2.1$ 
and the more recognizable one at $4.3$ are of unknown origin.

\subsection{$\xi$ non-integer}
\label{subsec:xinot}

\begin{figure}
  \centerline{\mbox{\epsfbox[0 0 270 397]{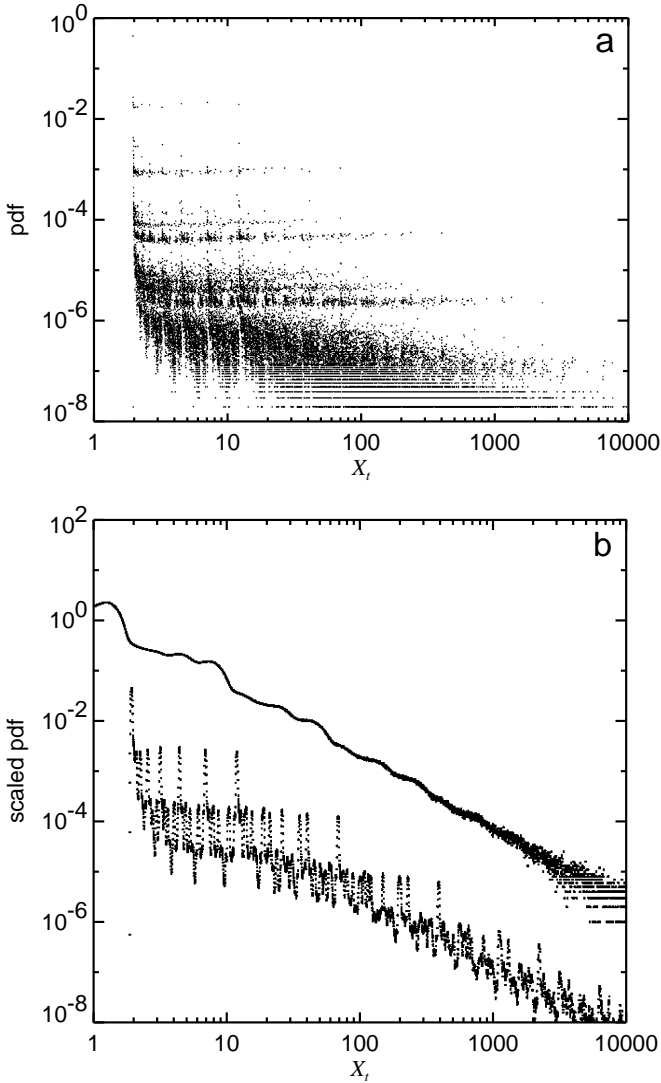}}}
  \caption{
           (a) The pdf for $a_t$ with two point distribution at 
           $a=2$, $\xi=2.5$, $p=0.95$, and $b_t=1$ ($10^8$ iterates, 
           $10^4$ equispaced bins per unit of $\log X_t$). (b) The 
           scaled pdfs for $a_t$ with two point distribution at $a=2$, 
           $\xi=2.5$, $p=0.95$, $\beta={15\over16}$ (lower trace, 
           $1\times{\rm pdf}$), and $\beta=0$ (upper trace, 
           $10^3\times{\rm pdf}$).  Both with $10^9$ iterates and 
           $10^3$ equispaced bins per unit of $\log X_t$.
          }
  \label{fig6}
\end{figure}

We have shown that for integer $\xi\ge5$ there will always exist 
some roots of (\ref{eq:zxip}) with non-zero imaginary and positive 
real parts.  This type of root structure is common for non-integer 
$\xi$.  If $\xi$ is irrational then an infinite number of distinct 
roots solves (\ref{eq:zxip}).  We select the slightly simpler case 
$\xi=2.5$ (with $a=2$ and $p=0.95$ as before).

In Fig.~\ref{fig6}~a, the pdf for $b_t=1$ is given for the $10^8$
first iterates of (\ref{fond}) with a binning density of $10^4$ 
points per decade.  Figure~\ref{fig6}~b shows the pdf for a uniformly 
distributed $b_t$ with $\beta={15\over 16}$ (lower trace) and the 
scaled pdf (upper trace, $10^3\times{\rm pdf}$) for $\beta = 0$, i.e. 
$b_t$ uniformly distributed between $0$ and $1$. These two pdfs use 
the first $10^9$ iterates of (\ref{fond}) with a log-equidistant 
binning of $10^3$ points per decade.

\begin{figure}
  \centerline{\mbox{\epsfbox[0 0 260 561]{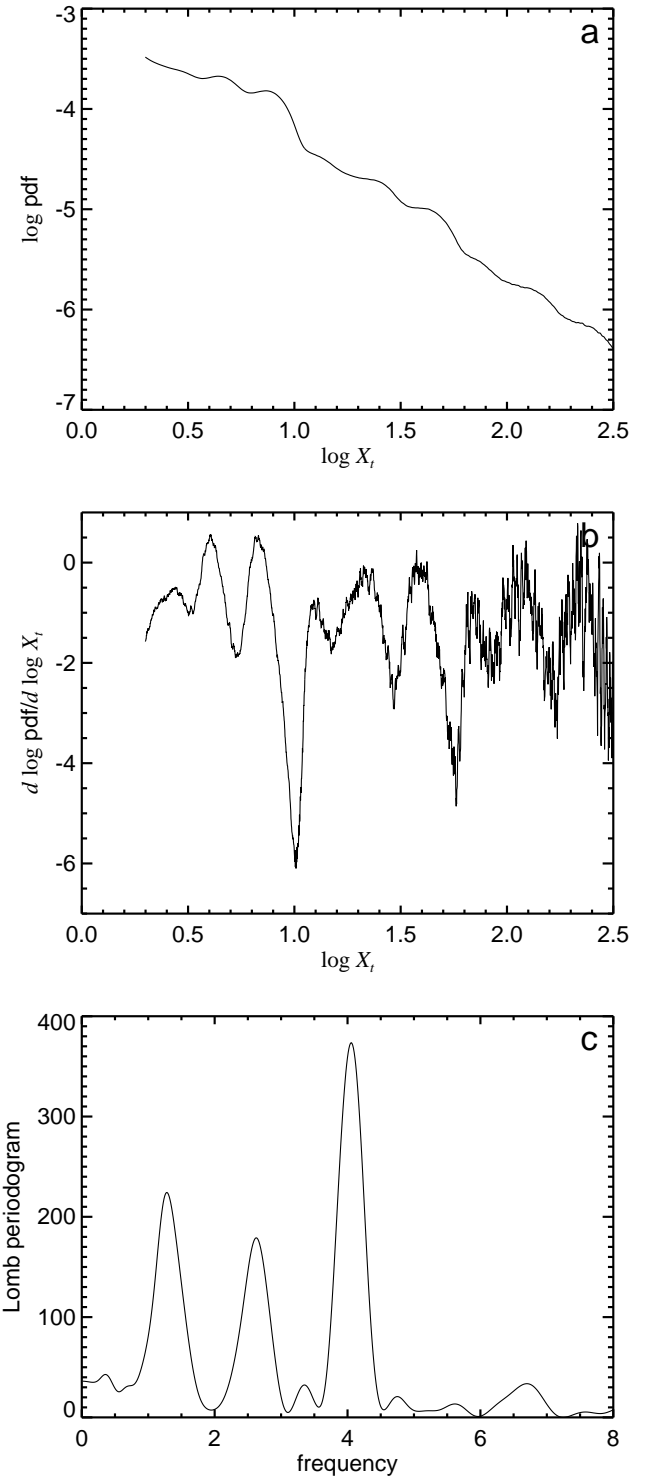}}}
  \caption{
           The pdf tail for $a_t$ with two point distribution at 
           $a=2$, $\xi=2.5$, $p=0.95$, and $b_t$ uniform with 
           $\beta=0$ ($10^9$ iterates, $10^3$ equispaced bins per 
           unit of $\log X_t$) (a).  (b) Its logarithmic derivative.  
           (c) The Lomb periodogram of (b).
          }
  \label{fig7}
\end{figure}

\begin{figure}
  \centerline{\mbox{\epsfbox[0 0 260 561]{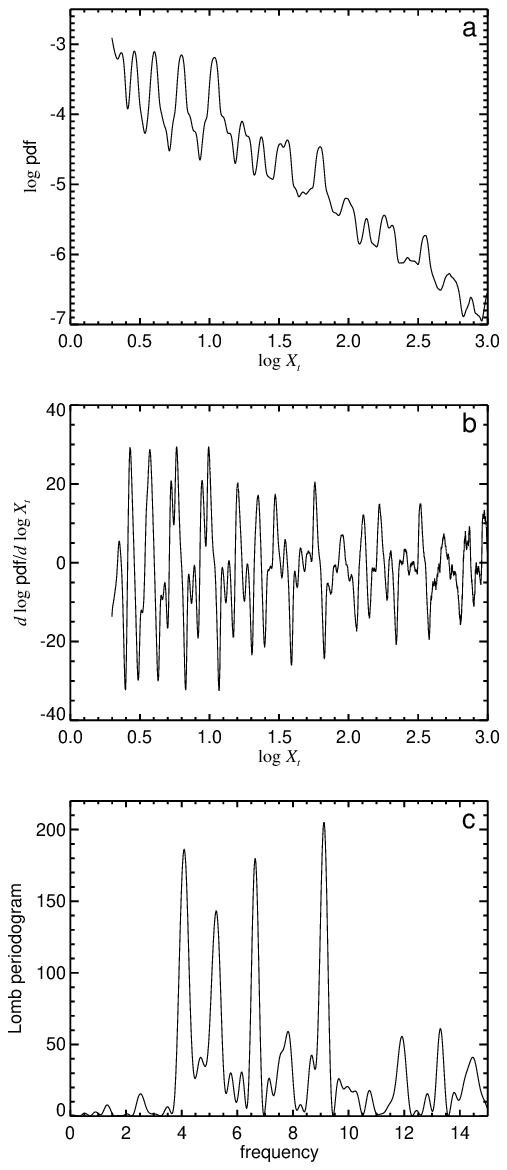}}}
  \caption{
           The pdf tail for $a_t$ with two point distribution at 
           $a=2$, $\xi=2.5$, $p=0.95$, and $b_t$ uniform with 
           $\beta={3\over4}$ ($10^9$ iterates, $10^3$ equispaced 
           bins per unit of $\log X_t$) (a).  (b) Its logarithmic 
           derivative.  (c) The Lomb periodogram of (b).
          }
  \label{fig8}
\end{figure}

\begin{figure}
  \centerline{\mbox{\epsfbox[0 0 260 561]{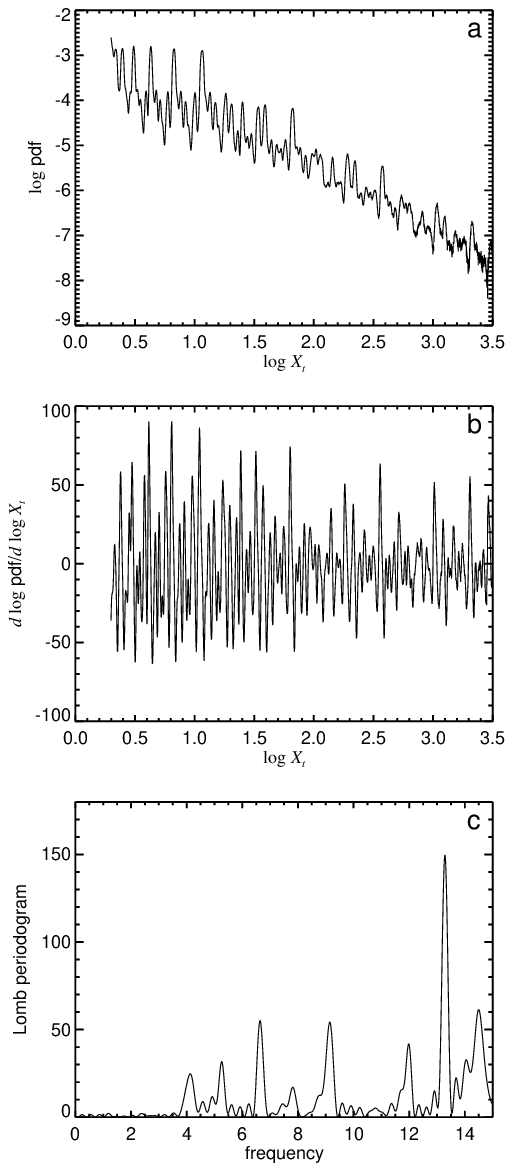}}}
  \caption{
           The pdf tail for $a_t$ with two point distribution at 
           $a=2$, $\xi=2.5$, $p=0.95$, and $b_t$ uniform with 
           $\beta={7\over8}$ ($10^9$ iterates, $10^3$ equispaced 
           bins per unit of $\log X_t$) (a).  (b) Its logarithmic 
           derivative.  (c) The Lomb periodogram of (b).
          }
  \label{fig9}
\end{figure}

As already pointed out in section~\ref{subsec:exis}, a continuous 
$b_t$ distribution seems to lead to a continuous pdf for $X$.  The 
analysis above neglected the influence of a varying $b_t$.   We 
resort to a limit consideration on a sequence of progressively 
thinned, uniform, $b_t$ distributions to match theory with the 
present simulation results.  We select the three cases $\beta=0, 
{3\over4}, {7\over8}$.  For each of these tail regions of the pdf 
(Fig.~\ref{fig7}~a, Fig.~\ref{fig8}~a, and Fig.~\ref{fig9}~a), the 
logarithmic derivative is computed (Fig.~\ref{fig7}~b, Fig.~\ref{fig8}~b, 
and Fig.~\ref{fig9}~b).  This gives a local estimate of the leading 
exponent of the power law tail of the pdf.  A constant value would
correspond to a pure power law.  Oscillations which are approximately 
periodic in $\ln X$ are the signatures of the log-periodicity.  This 
is confirmed by a spectral analysis given in Fig.~\ref{fig7}~c, 
Fig.~\ref{fig8}~c, and Fig.~\ref{fig9}~c of the signals shown in 
Fig.~\ref{fig7}~b, Fig.~\ref{fig8}~b, and Fig.~\ref{fig9}~b respectively, 
using the Lomb periodogram technique \cite{Numerci}.  We clearly 
identify a number of frequencies.  We compare these numerical results 
with a direct analytical determination of the roots of 
(\ref{mueq},\ref{eq:zxip}).  The complex $\mu$ solutions are sought 
where $z = a^{\mu} = e^{x+iy}$. These are the roots of
\be
   R(x,y;\xi,p) + i J(x,y;\xi,p) = 0 \;,
   \label{eq:RJez}
\ee
where
\ba
   R(x,y;\xi,p) &\equiv& (1-p) e^{(1+\xi)x}\cos((1+\xi)y) \nonumber \\
                & & \mbox{} -e^x \cos y + p \;,
                 \label{eq:Rdef}\\
   J(x,y;\xi,p) &\equiv& (1-p) e^{(1+\xi)x}\sin((1+\xi)y) \nonumber \\
                & & \mbox{} -e^x \sin y \;.
                 \label{eq:Jdef}
\ea
The zeroes of (\ref{eq:Rdef}) and (\ref{eq:Jdef}) define nodal curves
in the $x$--$y$  plane. The solutions $(x,y)$ are the intersections of
these nodal curves. When $\xi$ is rational $\xi=M/N$, where $M$ and $N$
are the smallest relative prime positive integers, we see that the set 
of solutions is periodic in the $y$-direction with a period $2\pi N$.
The two equations (\ref{eq:Rdef},\ref{eq:Jdef}) are converted into
\ba
   (1-p)e^{\xi x}&=& {\sin y \over
       \sin (1+\xi)y} \;,
   \label{aaz}\\
   e^x&=&p {\sin (1+\xi)y \over
     \sin \xi y} \;.
   \label{cccr}
\ea
This shows that the values of $x$ are bounded from above by a finite 
number.  An unbounded $x$ would from (\ref{cccr}) correspond to a 
vanishing $\sin\xi y$.  This would imply that $\xi y = n \pi$, for 
some integer $n$, and therefore that $\sin (1+\xi) y = (-1)^n \sin y$.  
Using (\ref{aaz}) we would obtain $(1-p)e^{\xi x} = 1$ ($n$ odd is not 
allowed) leading to a contradiction.  This implies that there is a 
maximum value for $\mu_R$.  Table~\ref{tabl1} gives the five solutions 
$(x_m,y_m)$, indexed by $m=1$ to $5$.  For a given $m$, we also extract 
the few first solutions obtained by $4 \pi$ periodic repetitions in the 
$y$ direction.  These solutions are indexed by an additional integer 
$n$ corresponding to the order of the $4\pi$ period.  We give these 
solutions in the $\mu_R$ and $\mu_I$ parameter space.

\vbox to 5.0cm
{
     \begin{table}
        \caption{
                 All roots $x$, $y$, and the first few $\mu=\mu_{R}+i 
                 \mu_{I}$ roots (here $a^{\mu} = e^{x+iy}$) of 
                 (\ref{eq:Rdef}, \ref{eq:Jdef}) for $a_t$ with two
                 point distribution at $a=2$, $\xi=2.5$, and $p=0.95$.
                 \label{tabl1}
                }
        \begin{tabular}{rccccc}
           \multicolumn{1}{c}{$m$} & \multicolumn{1}{c}{1} & 
              \multicolumn{1}{c}{2} & \multicolumn{1}{c}{3} & 
              \multicolumn{1}{c}{4} & \multicolumn{1}{c}{5} \\
           \tableline
           \multicolumn{1}{c}{$x(m)$} & 1.03 & 1.28 & 1.19 & 1.19 & 1.28 \\
           \multicolumn{1}{c}{$y(m)$} & 0.0 & 2.56 & 4.91 & 7.66 & 10.06 \\
           \multicolumn{1}{c}{$\mu_{R}(m)$} & 1.47 & 1.85 & 1.72 & 1.72 & 
                              1.85 \\
           \multicolumn{1}{c}{$\mu_{I}(m, n=0)$} & 0.0 & 3.69 & 7.08 & 
                              11.04 & 14.43 \\
           \multicolumn{1}{c}{$\mu_{I}(m, n=1)$} & 18.13 & 21.82 & 25.21 & 
                              29.18 & 32.57 \\
           \multicolumn{1}{c}{$\mu_{I}(m, n=2)$} & 36.26 & 39.95 & 45.34 & 
                              47.31 & 50.69 \\
           \multicolumn{1}{c}{$\mu_{I}(m, n=3)$} & 54.39 & 58.08 & 61.47 & 
                              65.43 & 68.82 
        \end{tabular}
     \end{table}
}

The appendix provides an approximate but quite accurate analytical
determination of the solutions found in this table, based on a 
perturbative scheme.  The pdf of $X_t$ is thus a sum of power laws 
overlayed by log-periodic oscillations of the type
\be
   P(X_t) = {C_{m,n} \over {X_t^{1+\mu_R(m)}}} \cos(\mu_I(m,n) \ln X_t) \;.
   \label{eq:pda1}
\ee
The leading power law behavior is given by the first $m=1$ real 
solution, which has the smallest $\mu=\mu_{R}(1) \approx 1.47$ 
(with $\mu_{I}(1,0)=0$).  The other solutions have larger $\mu_{R}$ 
and thus corresponds to subleading corrections.  We define the ``gap'' 
as the smallest difference between the real parts of the complex 
solutions to the first real solution.  This gap measures the strength 
of the subleading log-periodic corrections to the leading power law
behavior. In the present situation, all $\mu_{R}(m)$ take two values
$\approx 1.72$ and $\approx 1.85$, which are close to $\mu_{R}(1)$.  
The gap is approximately $0.25$.  This corresponds to strong corrections 
to the leading scaling for which the log-periodic oscillations are 
very visible.  Notice that, asymptotically, the oscillations disappear 
for $X_t \to \infty$, as $\mu_{R}(m>1) > \mu_{R}(1)$.  This effect is 
very weak in the present case, since the relative amplitude of the 
dominant log-periodic oscillations decays as $X_t^{-0.25}$.


\vbox to 9.6cm
{
     \begin{table}
        \caption{
                 The predicted and the observed frequencies (obtained by 
                 spectral analysis of Fig.~\ref{fig7}~c, Fig.~\ref{fig8}~c, 
                 and Fig.~\ref{fig9}~c).  The increased disorder on the 
                 variable $b_t$ is noted with the different $\beta$ subscript.
                 The first and the fith rows (the $F(M)$ labelled rows) 
                 contain the predicted frequencies.  The $F$-$\beta$ 
                 subscripted rows list the frequencies retrieved from the 
                 numerical realizations. The {\bf bold} type is used to 
                 indicate the frequency that gives the largest peak in the 
                 spectrum window, whereas the other well-defined peaks are 
                 given in standard type. The parenthesis enclosed values
                 correspond to spectrum peaks that barely are above the 
                 noise level of the Lomb periodgram.
                 \label{tabl2}
                }
        \begin{tabular}{rccccc}
           \multicolumn{1}{c}{$M$} & \multicolumn{1}{c}{1} & 
              \multicolumn{1}{c}{2} & \multicolumn{1}{c}{3} & 
              \multicolumn{1}{c}{4} & \multicolumn{1}{c}{5} \\
           \tableline
           \multicolumn{1}{c}{$F(M)$} & 1.35 & 2.60 & 4.05 & 5.29 & 6.64 \\
           \multicolumn{1}{c}{$F_{\beta=7/8}(M)$} & - & - & 4.13 & 5.25 & 
                              6.63 \\
           \multicolumn{1}{c}{$F_{\beta=3/4}(M)$} & (1.33) & (2.53) & 4.08 & 
                              5.25 & 6.65 \\
           \multicolumn{1}{c}{$F_{\beta=0}(M)$} & 1.31 & 2.64 & $\bf{4.08}$ & 
                              - & (6.69) \\
           \tableline
           \multicolumn{1}{c}{$M$} & \multicolumn{1}{c}{6} & 
              \multicolumn{1}{c}{7} & \multicolumn{1}{c}{8} & 
              \multicolumn{1}{c}{9} & \multicolumn{1}{c}{10} \\
           \tableline
           \multicolumn{1}{c}{$F(M)$} & 8.00 & 9.24 & 10.69 & 11.93 & 13.29 \\
           \multicolumn{1}{c}{$F_{\beta=7/8}(M)$} & (7.83) & 9.13 & - & 
                              11.96 & $\bf{13.25}$ \\
           \multicolumn{1}{c}{$F_{\beta=3/4}(M)$} & (7.84) & $\bf{9.12}$ & 
                              10.77 & 11.92 & 13.30 \\
           \multicolumn{1}{c}{$F_{\beta=0}(M)$} & - & - & - & - & - \\
        \end{tabular}
     \end{table}
}

Table~\ref{tabl2} gives the observed frequencies obtained by the
spectral analysis of Fig.~\ref{fig7}~c, Fig.~\ref{fig8}~c, and
Fig.~\ref{fig9}~c and compare them with the predicted values. 
This contains the different cases with increased disorder on the 
variable $b_t$.  Here a slight generalization of frequency is used, 
$f(m,n) \equiv {\mu_{I}(m,n)\over 2\pi}$ and 
$F(M) \equiv f(M+1-5n,n=[{M\over5}])$.  Increasing the disorder 
in the variable $b_t$ increases the noise level and progressively 
washes out the higher frequencies.

\section{TWO LEVEL ``STAIRCASE'' DISTRIBUTION}
\label{sec:twol}

\begin{figure}
  \centerline{\mbox{\epsfbox[0 0 260 182]{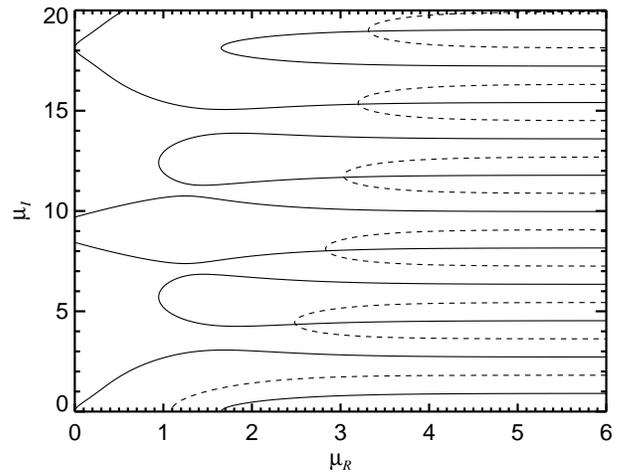}}}
  \caption{
           A portion of the complex $\mu$ plane with roots 
           $\mu=\mu_{R}+i\mu_{I}$ of (\ref{eq:adl5}) (for 
           $a_t$ with a two level staircase distribution at 
           $a=2$, $\xi=2.5$, $p=0.95$) as intersections between 
           its real part (solid lines) and imaginary part 
           (dashed lines).
          }
  \label{fig10}
\end{figure}

We now study a situation with a much larger disorder where the 
multiplicative factors $a_t$ are selected from a broad, continuous, 
distribution.  To minimize the number of control parameters for the 
pdf's, we use distributions which are constant by parts. In the next 
section we will examine the uniform distribution.  Here, we divide 
the interval $[{1 \over a}, a^{\xi}]$ into two sub-intervals 
$[{1 \over a},1]$ and $[1, a^{\xi}]$ having different weights $p$ 
and $1-p$ respectively.  The idea is to allow for a different weight 
of the damping versus amplificating processes and examine the 
consequence on the amplitude of the log-periodic structures.  
This choice corresponds to the following pdf for $a_t$\,:
\ba
   P_{a_t}(a_t) &=& {{p}\over{1-1/a}}(\Theta(a_t-1/a) - \Theta(a_t-1))
      \nonumber \\
                & & \mbox{} +{{1-p}\over{a^{\xi}-1}}(\Theta(a_t-1) -
            \Theta(a_t-a^{\xi})) \;, 
      \label{eq:adl1}
\ea
where $\Theta$ is the Heaviside function.  The stationarity condition 
(\ref{eq:lna0}) that $\langle \ln a_t \rangle <0$ reads
\be
   \check{p}(a,\xi) < p < 1 \;,
   \label{eq:adl2}
\ee
where
\ba
   \check{p}(a,\xi)&\equiv&{{a^{\xi}(\xi\ln a-1)+1}\over
        {a^{\xi}(\xi\ln a-1)+1+\delta(a,\xi)}} \nonumber \\
   \mbox{with}\quad && \delta(a,\xi)={a^{\xi}-1\over{a-1}}(a-1-\ln a) \;.
   \label{eq:adl3}
\ea
The integral equation (\ref{eq:pdf2}) is now
\be
   P(X)={{a p}\over{a-1}}\int_{1/a}^{1} {P({X\over a_{t}})\over a_{t}} d a_{t}
        + {{1-p}\over{a^{\xi}-1}}\int_{1}^{a^{\xi}}
          {P({X \over a_{t}})\over a_{t}} d a_{t} \;.
   \label{eq:adl4}
\ee
This equation has a power law solution for large $X$ if the exponent 
$\mu$ satisfies (\ref{mueq}), leading to
\be
   \mu+1={{a p}\over{a-1}}(1-a^{-(\mu+1)})
         + {{1-p}\over{a^{\xi}-1}}(a^{\xi(\mu+1)}-1) \;.
   \label{eq:adl5}
\ee
Assuming a complex solution $\mu=\mu_{R}+i \mu_{I}$ splits equation
(\ref{eq:adl5}) into
\ba
   \mu_{R} + 1 &=& {{a p}\over{a-1}}(1-a^{-(\mu_{R}+1)}\cos({\mu_{I}\ln a}))
      \nonumber \\
               & & \mbox{} +
      {{1-p}\over{a^{\xi}-1}}(a^{\xi(\mu_{R}+1)}
         \cos({\xi\mu_{I}\ln a})-1) \;,
   \label{eq:ad6r}\\
   \mu_{I} &=& {{a p}\over{a-1}}a^{-(\mu_{R}+1)}\sin({\mu_{I}\ln a}) 
      \nonumber \\
           & & \mbox{} +
      {{1-p}\over{a^{\xi}-1}}a^{\xi(\mu_{R}+1)}
         \sin({\xi\mu_{I}\ln a}) \;.
   \label{eq:ad6i}
\ea
To allow for a comparison with the previous case, we keep the same 
parameters $a=2$, $\xi = 2.5$ and $p=0.95$ as before.  The solutions 
of these equations are graphically represented as the intersections 
of the  continuous and dashed lines in Fig.~\ref{fig10} (here 
$\check{p}(2.0,2.5)\approx 0.78266$).  Table~\ref{tabl3} lists the 
smallest roots and their corresponding log-periodic frequencies 
($f(m)\equiv{{\mu_{I}(m)}\over{2 \pi}}$).

\begin{figure}
  \centerline{\mbox{\epsfbox[0 0 260 182]{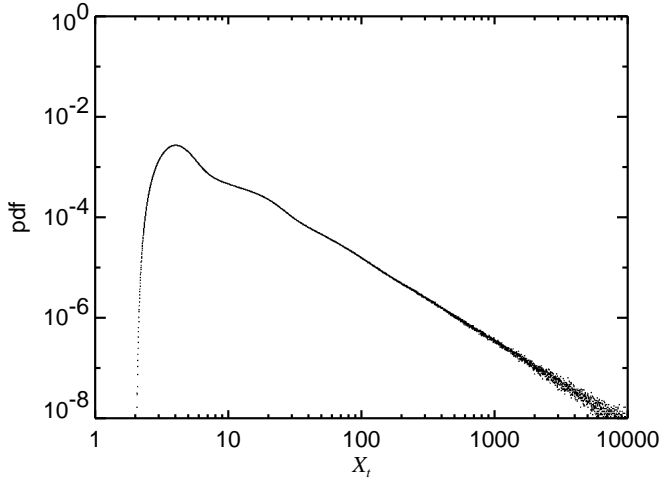}}}
  \caption{ 
           The pdf for the case where the pdf of $a_t$ has a level 
           staircase structure at $a=2$, $\xi=2.5$, $p=0.95$, with 
           $b_t=1$ ($10^9$ iterates, $10^3$ equispaced bins per unit 
           of $\log X_t$).
          }
  \label{fig11}
\end{figure}

\vbox to 3.5cm
{
     \begin{table}
        \caption{
                 The first few $\mu=\mu_{R}+i \mu_{I}$ roots of 
                 (\ref{eq:ad6r}, \ref{eq:ad6i}) and the predicted 
                 log-periodic frequencies for $a_t$ with a two level 
                 staircase distribution at $a=2$, $\xi=2.5$, and $p=0.95$.
                 \label{tabl3}
                }
        \begin{tabular}{rcccccc}
           \multicolumn{1}{c}{$m$} & \multicolumn{1}{c}{1} & 
              \multicolumn{1}{c}{2} & \multicolumn{1}{c}{3} & 
              \multicolumn{1}{c}{4} & \multicolumn{1}{c}{5} &
              \multicolumn{1}{c}{6} \\
           \tableline
           \multicolumn{1}{c}{$\mu_{R}(m)$} & 1.6535 & 2.4918 & 2.8418 &
                              3.0389 & 3.2015 & 3.3179 \\
           \multicolumn{1}{c}{$\mu_{I}(m)$} & 0.0000 & 4.3475 & 8.0152 &
                              11.6794 & 15.3255 & 18.9598 \\
           \multicolumn{1}{c}{$f(m)$} & 0.0000 & 1.5932 & 2.9373 &
                              4.2801 & 5.6163 & 6.9482 \\
        \end{tabular}
     \end{table}
}

\begin{figure}
  \centerline{\mbox{\epsfbox[0 0 260 561]{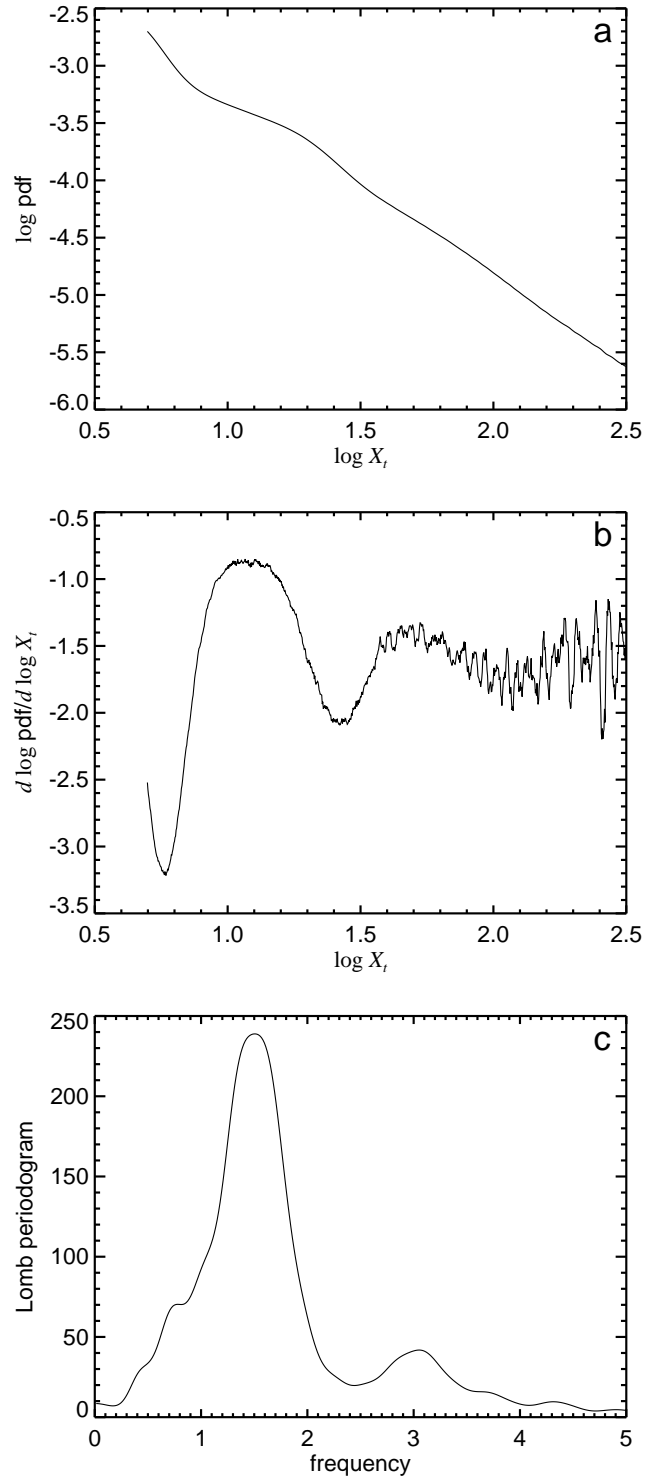}}}
  \caption{
           (a) The Pdf tail for $a_t$ with two point distribution at 
           $a=2$, $\xi=2.5$, $p=0.95$, and $b_t=1$ ($10^9$ iterates, 
           $10^3$ equispaced bins per unit of $\log X_t$).  (b) The 
           logarithmic derivative of (a). (c) The Lomb periodogram of (b).
          }
  \label{fig12}
\end{figure}

An important difference with the previous two-point pdf is that 
now the gap value is $\mu_{R}(2) - \mu_{R}(1) \simeq 0.84$, which 
is about three times larger than before.  This means that the 
log-periodic structure are smaller and decay faster for large $X$.  
They are still quite visible as found in Fig.~\ref{fig11}, where 
we can observe the undulation of $P(X)$'s tail. The results 
obtained for the various $b_t$ distributions (from a non-random 
$b_{t}=1$ to a uniform $b_t$ with $\beta=0$) are essentially the 
same for $X \geq 3$. The only difference is that a larger disorder 
in $b_t$ allows for an exploration of the interval closer to $0$.
The Lomb power spectrum analysis is presented in Fig.~\ref{fig12}.
The fundamental frequency, $f(2)\approx 1.59$, is visible in all 
the simulations and more clearly in the spectral analysis where 
a strong peak appears in the Lomb power spectrum.  The next higher 
frequency, $f(3)\approx 3.0$, is the only one which can be detected 
as the disorder in $b_t$ increases. All higher frequencies are lost 
in the noise.  The reason for this is clear.  The relative amplitude 
of a given frequency $f(m)$ is quantified by $\mu_{R}(m) - \mu_{R}(1)$. 
For the second frequency, we have $\mu_{R}(3) - \mu_{R}(1) = 1.19$. 
For the third frequency, we have $\mu_{R}(4) - \mu_{R}(1) = 1.39$ and 
so on. It seems that a difference $\mu_{R}(m) - \mu_{R}(1)$ of the 
order or less than $1$ is necessary for the clear detection of 
log-periodicity.  Intuitively, this ensures that the amplitude of 
the oscillations does not decay more than by a factor $100$ over 
two decades. We notice that a similar gap about $1$ was found in the
analysis of the log-periodic structure of DLA clusters \cite{DLA}. 
The present analysis rationalizes why we have been able to detect these 
structures in this case.

Table~\ref{tabl4} makes the comparison between the predicted and observed
frequencies.

\narrowtext
\vbox to 6.6cm
{
     \begin{table}
        \caption{
                 Predicted and observed frequencies for $a_t$ with a 
                 two level staircase distribution at $a=2$, $\xi=2.5$, 
                 $p=0.95$, and two different choices for $b_t$ distribution. 
                 The first row contains the predicted frequencies. 
                 Subsequent rows are the frequencies retrieved from the 
                 numerical simulations. The {\bf bold} emphasis indicates 
                 the frequency that gives the largest peak in the spectrum.
                 The other well-defined peaks are written in normal format, 
                 while numbers inside parenthesis correspond to peaks in 
                 the spectrum that are barely above the noise level of 
                 the Lomb periodgram.
                 \label{tabl4}
                }
        \begin{tabular}{rccc}
           \multicolumn{1}{c}{$m$} & \multicolumn{1}{c}{1} & 
              \multicolumn{1}{c}{2} & \multicolumn{1}{c}{3} \\
           \tableline
           \multicolumn{1}{c}{$f(m+1)$} & 1.59 & 2.94 & 4.28 \\
           \multicolumn{1}{c}{$f_{b_t=1}(m)$} & $\bf{1.51}$ & 3.07 & (4.31) \\
           \multicolumn{1}{c}{$f_{\beta=0}(m)$} & $\bf{1.51}$ & 3.00 & (4.50) \\
        \end{tabular}
     \end{table}
}

We conclude that a log-periodic structure of the tail of the $X_t$'s pdf
is present for the smeared out two level staircase distribution, although
its amplitude is weakened as compared to the previous two point distribution
case.  This was expected from the theoretical analysis of the influence of
disorder \cite{SS,DLA}.  The important aspect of our result is that the
log-periodicity, and the prefered scaling ratios $\lambda$, can no more be
associated to a specifically chosen amplification factor, as for the previous
two-point distribution.  Notwithstanding the presence of a large disorder, a
discrete set of effective scaling factors are selected.  It is amazing to us
how strong is this effect and how relatively weak is the influence of the
disorder.

\section{UNIFORM DISTRIBUTION}
\label{sec:unif}

\begin{figure}
  \centerline{\mbox{\epsfbox[0 0 260 182]{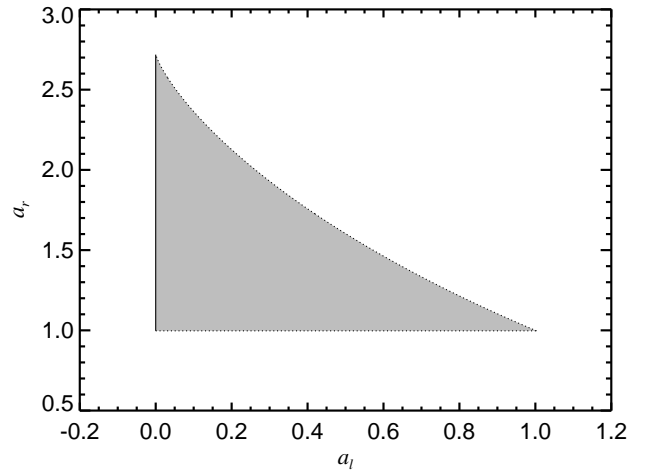}}}
  \caption{
           Allowed $a_{l}$--$a_{r}$ domain for $a_t$ with uniform 
           distribution.  The dotted boundary depicts the strict 
           inequalities given in (\ref{eq:adu3}).
          }
  \label{fig13}
\end{figure}

This last section deals with the effects of a very strong disorder on 
$a_t$.  To attain this goal, we consider a uniform $a_t$-distribution
\be 
   P_{a_t}(a_t) = {{\Theta(a_t-a_\ell) - \Theta(a_t-a_r))}
                      \over{a_r-a_\ell}} \;,             
   \label{eq:adu1}
\ee
where
\ba
   0 &\le& a_\ell < 1 \;,
     \label{eq:adu2}\\
   1 &<& a_r < \hat a_r(a_\ell) \;.
     \label{eq:adu3}
\ea
$\hat a_r(a_\ell)$ is such that (\ref{eq:lna0}) is obeyed and is the 
solution of
\be
   \hat a_r(a_\ell) \ln \hat a_r(a_\ell) - \hat a_r(a_\ell) =
       a_\ell \ln a_\ell - a_\ell \;.
   \label{eq:adu4}
\ee
Figure~\ref{fig13} shows the allowed $a_\ell$--$a_r$ region.

The largest width compatible with (\ref{eq:lna0}) corresponds to 
$a_\ell=0$ and $a_r=e$.  When $a_\ell$ is made larger, the maximum 
value of $a_r$ progressively decays towards $1$.

The integral equation (\ref{eq:pdf2}) is
\be
   P(X)={{1}\over{a_r-a_\ell}}\int_{a_\ell}^{a_r}
               {{P({X \over a_{t}})}\over a_{t}}d a_{t} \;.
   \label{eq:adu5}
\ee
The tail of $P(X)$ takes the form of a power law if the exponent 
$\mu$ is the solution of
\be
   (\mu-1)(a_r-a_\ell)=a_r^{\mu+1}-a_\ell^{\mu+1} \;.
   \label{eq:adu6}
\ee
With $\mu=\mu_{R}+i \mu_{I}$, we get
\ba
   (\mu_{R}+1)(a_r-a_\ell) &=& a_r^{\mu_{R}+1}\cos({\mu_{I}\ln a_r}) 
       \nonumber \\
                           & & \mbox{} -
   a_\ell^{\mu_{R}+1}\cos({\mu_{I}\ln a_\ell}) \;,        
   \label{eq:adu7}\\
   \mu_{I}(a_r-a_\ell) &=& a_r^{\mu_{R}+1}\sin({\mu_{I}\ln a_r}) \nonumber \\
                       & & \mbox{} -
       a_\ell^{\mu_{R}+1}\sin({\mu_{I}\ln a_\ell}) \;.
   \label{eq:adu8}
\ea

We select $a_\ell=0.001$ and $a_r=1.9$.  The solutions with the 
smallest, positive, real parts are given in the table~\ref{tabl5}, 
together with their corresponding log-frequencies $f$.

\vbox to 3.8cm
{
     \begin{table}
        \caption{
                 The first few $\mu=\mu_{R}+i \mu_{I}$ roots of 
                 (\ref{eq:adu7}, \ref{eq:adu8}) and the predicted 
                 log-periodic frequencies for the uniformly distributed 
                 $a_t$ with $a_\ell=0.001$ and $a_r=1.9$.
                 \label{tabl5}
                }
        \begin{tabular}{rccccccc}
           \multicolumn{1}{c}{$m$} & \multicolumn{1}{c}{1} & 
              \multicolumn{1}{c}{2} & \multicolumn{1}{c}{3} & 
              \multicolumn{1}{c}{4} & \multicolumn{1}{c}{5} \\
           \tableline
           \multicolumn{1}{c}{$\mu_{R}(m)$} & 1.2667 & 3.9414 & 4.8349 & 
                              5.3984 & 5.8116 \\
           \multicolumn{1}{c}{$\mu_{I}(m)$} & 0.0000 & 11.6095 & 21.6147 & 
                              31.5024 & 41.3494 \\
           \multicolumn{1}{c}{$f(m)$} & 0.0000 & 4.2545 & 7.9211 & 11.5446 & 
                              15.1532 \\
        \end{tabular}
     \end{table}
}

\begin{figure}
  \centerline{\mbox{\epsfbox[0 0 260 182]{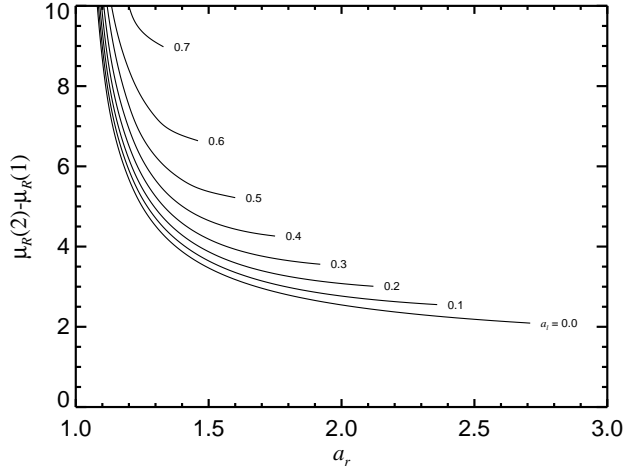}}}
  \caption{
           The gap value, $\mu_{R}(2)-\mu_{R}(1)$, as a function of 
           $a_{r}$ for different choices of $a_{l}$.\hfill
          }
  \label{fig14}
\end{figure}

The most striking feature to note is the large gap value,
$\mu_{R}(2)-\mu_{R}(1) \simeq 2.67$.  These differences 
increase rapidly with the order $m$ of the solution. This 
implies that the oscillations must be extremely weak and 
severely dampened.  In Fig.~\ref{fig14} we explore the 
dependence of the gap as a function of the parameters
$a_\ell$ and $a_r$ of the model. If $a_\ell=0$ and $a_r=2.71$, 
we find $\mu_{R}(2)-\mu_{R}(1) \simeq 2.1$, but $\mu_{R}(1)=0.006$ 
is very small.  When $a_\ell=0$ and $a_r=2.0$, we find 
$\mu_{R}(2)-\mu_{R}(1)=2.545$ with $\mu_{R}(1)=1.0$.

\begin{figure}
  \centerline{\mbox{\epsfbox[0 0 260 182]{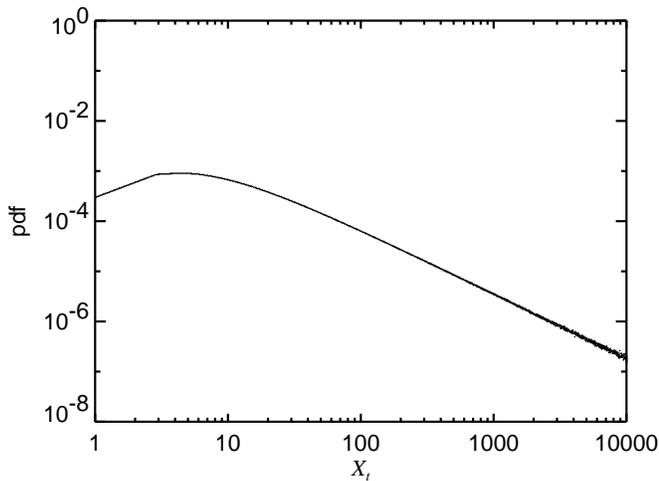}}}
  \caption{
           The pdf for $a_t$ with uniform distribution between 
           $a_{l}=0.001$ and $a_{r}=1.9$; $b_t=1$ ($10^9$ iterates, 
           $10^3$ equispaced bins per unit of $\log X_t$).
          }
  \label{fig15}
\end{figure}

The situation does not improve if we take $a_\ell \to 1^-$ and $a_r \to
1^+$ (while keeping the stationarity condition (\ref{eq:lna0})). Notice
that we need $a_r > 1$ in order to get a solution for $\mu$, {\it i.e} to
get a power law pdf for $X_t$. This stems from the fundamental fact that
the power law pdf results from intermittent amplifications.  In summary,
the log-periodic oscillations are present theoretically but are very
difficult to measure and quantify. The pdf for the $b_t=1$ case is shown
in Fig.~\ref{fig15}.

\begin{figure}
  \centerline{\mbox{\epsfbox[0 0 260 369]{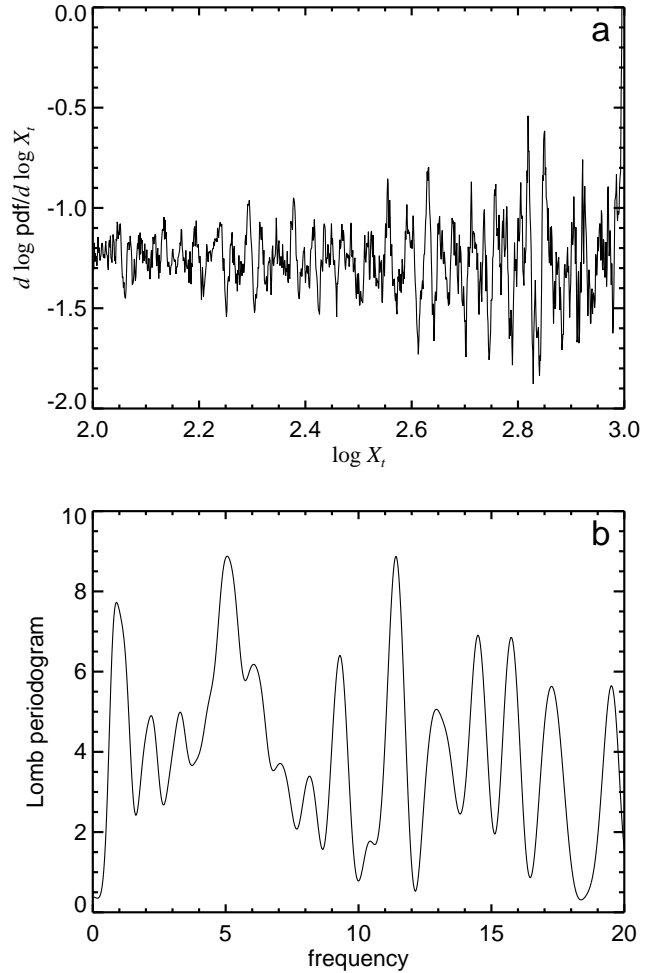}}}
  \caption{
           (a) The logarithmic derivative of the pdf of $X$ for $a_t$ 
           with a uniform distribution between $a_{l}=0.001$ and 
           $a_{r}=1.9$, $b_t=1$ ($10^9$ iterates, $10^3$ equispaced 
           bins per unit of $\log X_t$). (b) The Lomb periodogram of (a).
          }
  \label{fig16}
\end{figure}

A segment of its tail is analyzed with the same procedure as was applied
to the previous two $a_t$-distribution families (Fig.~\ref{fig16}).

Only $f(4)$ seems to be recognizable among all the peaks in the Lomb
periodogram but this seems even far-fetched as the signal is within 
the noise level.  In conclusion, this uniform case corresponds to a 
large gap and the log-periodic structures, that are present in theory, 
are not clearly visible.

\section{CONCLUDING REMARKS}
\label{sec:conc}

We started our analysis by considering the intermittent multiplicative
processes of a simple binomial $a_t$ distribution.  Not surprisingly,
strong log-periodic corrections to the main power law probability density
function have been found for the random affine map.  The disorder in the
additive constant smooths out the higher frequencies but does not dampen
out the smallest log-periodic frequencies. We then analyzed situations
with increasing disorder in the multiplicative terms.  Instead of going 
to the weak disorder regime with two broadened peaks, we analyzed a pdf 
of multiplicative factors which consists of a two-step staircase.  In 
this already large disorder regime, we have found that the log-periodic 
structure of the tail of the $X_t$'s pdf is present for the smeared out 
two level staircase distribution, although it is  weakened in its amplitude 
as compared to the two point distribution case.  The most important aspect 
of our results is that the log-periodicity, and the prefered scaling ratios
$\lambda$, can no more be associated to a specifically chosen amplification
factor, as for the two-point distribution. Notwithstanding the presence of 
a large disorder, a discrete set of effective scaling factors are selected. 
The ``gap'', defined as the difference between the smallest exponent real 
part and the real solution, controls the strength of the log-periodicity. 
We have been able to determine that the gap must be of the order or less 
than $1$ in order for the log-periodicity to be strong.  Larger gaps still 
lead to visible effects but the analysis must then be very precise and the 
noise level very low. This is the situation found for a uniform distribution 
of multiplicative factors.  In summary, we have shown that log-periodicity 
remains a significant effect even in the presence of significant disorder.

\section{APPENDIX\,: DETERMINATION OF THE EXPONENTS FOR THE TWO-POINT
DISTRIBUTION USING A PERTURBATIVE ANALYSIS}
\label{sec:appn}

The power law structure of the $P(X)$ pdf characterizes rare excursions of
$X_t$ to large values. These large values are reached by repeated occurrence
of the amplifying multiplicative factor $a^{\xi}$. This motivates us to make
the approximation of neglecting the first ``damping'' term in the r.h.s. of
(\ref{eq:pft1}),
\be
   P(X) \approx (1-p) a^{-\xi} P(a^{-\xi} X) \;.
   \label{eq:pft2}
\ee
This functional equation is simpler to handle.  It also has a form reminiscent
of the renormalization group equation that Feigenbaum used in his analysis of
a bifurcation sequence of the logistic equation \cite{Feigen}. Analoguous
equations has also been discussed in \cite{Anifrani,SorSam,SSS,SS}. Assuming
a power law form for $P(X)$ ($P(X)={A \over X^{1+\mu}}$) provides for the
$\mu$\ equation\,:
\be
   (1-p)a^{\mu \xi}=1 \;.
   \label{eqap:sec}
\ee
With the notation $z=e^{x+iy}=a^{\mu_{R}+i \mu_{I}}$, we obtain
\ba
   x&=&-{{1}\over{\xi}}\ln(1-p)\quad\mbox{and}\quad
         \mu_{R}=-{{\ln(1-p)}\over{\xi\ln a }}
   \label{eq:rgr1}\\
   y&=&{{2 k \pi}\over{\xi}}\quad\mbox{and}\quad
         \mu_{I}={{2 \pi}\over{\ln \lambda_{k}}} \;, \nonumber \\
    & & \mbox{} \quad\mbox{where}\quad
         \lambda_{k}\equiv a^{\xi/k},\quad k \quad \mbox{ integer} \;.
   \label{eq:rgi1}
\ea
An approximation of the imaginary part of the roots of (\ref{eq:RJez})
therefore is
\be
   y={2 k \pi \over \xi} \;.
   \label{eq:yzoz}
\ee
This agrees fairly well with the exact computed roots shown in
table~\ref{tabl1}.  We can improve on this estimation by inserting 
the parametrization
\be
   y={2 k \pi \over \xi} + \epsilon \;,
   \label{eq:yzoo}
\ee
in the full set of equations (\ref{eq:Rdef}) and (\ref{eq:Jdef}). This
yields the following system of two equations of the two unknown $x$ and
$\epsilon$\,:
\ba
   & &((1-p)e^{(1+\xi)x}\cos\epsilon(1+\xi)-e^{x}\cos\epsilon)
       \cos {{2k\pi}\over\xi}
    \nonumber \\
   & &\mbox{ }-((1-p) e^{(1+\xi)x}\sin\epsilon(1+\xi) \nonumber \\
   & & \mbox{\hspace{1cm}}\;\; -e^{x}\sin\epsilon)
     \sin {{2k\pi}\over \xi} + p = 0 \;,
   \label{eq:rgr2}\\
   & &((1-p)e^{\xi x}\cos\epsilon(1+\xi)-\cos\epsilon)\sin {{2k\pi}\over \xi}
     \nonumber \\
   & &\mbox{ }\;\;+((1-p) e^{\xi x}\sin\epsilon(1+\xi) \nonumber \\
   & & \mbox{\hspace{1cm}} -\sin\epsilon)
      \cos {{2k\pi}\over \xi} = 0 \;.
   \label{eq:rgi2}
\ea
Assuming $\epsilon(1+\xi)$ to be ``small'', we expand the trigonometric
functions to first order in $\epsilon$\,:
\ba
   &(&(1-p)e^{(1+\xi)x}-e^{x})\cos {2k\pi/\xi}-
      \epsilon((1-p)(1+\xi)e^{(1+\xi)x} \nonumber \\
   & & \mbox{\hspace{1cm}} -e^{x})\sin {2k\pi/\xi} + p = 0 \;,
   \label{eq:rgr3}\\
   &(&(1-p)e^{\xi x}-1)\sin {2k\pi/\xi} + \epsilon((1-p)(1+\xi)e^{\xi x} 
     \nonumber \\
   & & \mbox{\hspace{1cm}} -1)\cos {2k\pi/\xi} = 0 \;.
   \label{eq:rgi3}
\ea
Eliminating $\epsilon$ between the two preceding equations, we get an
equation in the sole variable $x$\,:
\be
   (1-p)e^{(1+\xi)x}-e^{x}+p \cos{2k\pi/\xi}=0 \;.
   \label{eq:eps2}
\ee
There is a unique solution in $x$ for each $k$.  Knowing $x$, we then get
$\epsilon$ from
\be
   \epsilon=-{{(1-p)e^{\xi x}-1}\over{(1-p)(1+\xi)e^{\xi x}-1}}\,
          \tan {2k\pi \over \xi} \;.
   \label{eq:eps1}
\ee
Table~\ref{tabl6} compares these solutions to the exact ones, in the case
when $a=2$, $\xi=2.5$, $p=0.95$.

\vbox to 4.1cm
{
     \begin{table}
        \caption{
                 The appoximate (according to (\ref{eq:eps2}, 
                 \ref{eq:eps1}, \ref{eq:yzoo})) and the exact 
                 (from (\ref{eq:Rdef}, \ref{eq:Jdef})) roots 
                 for $a_t$ with two point distribution at $a=2$, 
                 $\xi=2.5$, and $p=0.95$.
                 \label{tabl6}
                }
        \begin{tabular}{rcccccc}
           \multicolumn{1}{c}{$k$} & \multicolumn{1}{c}{0} & 
              \multicolumn{1}{c}{1} & \multicolumn{1}{c}{2} & 
              \multicolumn{1}{c}{3} & \multicolumn{1}{c}{4} & 
              \multicolumn{1}{c}{5} \\
           \tableline
           \multicolumn{1}{c}{$x_{apprx}(k)$} & 1.0333 & 1.2760 & 1.1597 & 
                              1.1597 & 1.2760 & 1.0333 \\
           \multicolumn{1}{c}{$x_{exact}(k)$} & 1.0333 & 1.2808 & 1.1922 & 
                              1.1922 & 1.2808 & 1.0333 \\
           \multicolumn{1}{c}{$\epsilon(k)$} & 0.0000 & 0.0480 & -0.1301 & 
                              0.1301 & -0.0480 & 0.0000 \\
           \multicolumn{1}{c}{$y_{apprx}(k)$} & 0.0000 & 2.5612 & 4.8965 & 
                              7.6699 & 10.0051 & 12.5664 \\
           \multicolumn{1}{c}{$y_{exact}(k)$} & 0.0000 & 2.5605 & 4.9101 & 
                              7.6563 & 10.0058 & 12.5664 \\
        \end{tabular}
     \end{table}
}

\end{multicols}
\end{document}